\def\bd{\begin{displaymath}}\def\ed{\end{displaymath}}
\def\be{\begin{equation}}\def\ee{\end{equation}}
\def\bea{\begin{eqnarray}}\def\eea{\end{eqnarray}}
\def\ba{\begin{array}}\def\ea{\end{array}}
\def\lb{\label}

\def\d{\delta}
\def\g{\gamma}
\def\k{\kappa}\def\m{\mu}\def\t{\tau}
\def\x{\xi}

\def\D{\Delta}\def\G{\Gamma}

\def\id{\equiv}\def\ha{{1\over 2}}

\def\trans{transformations }
\def\tran{transformation }\def\coo{coordinates }

\def\pb{Poisson brackets }

\def\min{Minkowski }
\def\poi{Poincar\'e }

\def\PL#1{Phys.\ Lett.\ {\bf#1}}
\def\PRL#1{Phys.\ Rev.\ Lett.\ {\bf#1}}
\def\PR#1{Phys.\ Rev.\ {\bf#1}}\def\CQG#1{Class.\ Quantum Grav.\ {\bf#1}}

 \def\IJMP#1{Int.\ J. Mod.\ Phys.\ {\bf #1}}
\def\MPL#1{Mod.\ Phys.\ Lett.\ {\bf #1}} 

\def\AoP#1{Ann.\ Phys.\ {\bf#1}}
\def\grq#1{{\tt gr-qc/#1}}\def\hep#1{{\tt hep-th/#1}}

\def\den{\left(1-{p_0\over\k}\right)}\def\dem{\left(1-{M\over\k}\right)}
\def\emm{e^{-2M/\k}}

\documentclass[12pt]{article}
\begin{document}

\begin{titlepage}
\vspace{.3cm}
\begin{center}
\renewcommand{\thefootnote}{\fnsymbol{footnote}}
{\Large \bf Lifetime of flying particles in
canonical Doubly Special Relativity}
\vspace*{2.4cm}

{\large \bf {S.~Mignemi\footnote{email: smignemi@unica.it}}}\\
\renewcommand{\thefootnote}{\arabic{footnote}}
\setcounter{footnote}{0}
\vspace*{0.8cm}
{\small
Dipartimento di Matematica, Universit\`a di Cagliari,\\
Viale Merello 92, 09123 Cagliari, Italy\\
\vspace*{0.4cm}
 INFN, Sezione di Cagliari\\}
\end{center}
\vspace*{2.4cm}
\centerline{\bf Abstract}
\vspace*{0.8cm}
\noindent
We discuss the corrections to the lifetime of unstable elementary particles
in some models of doubly special relativity. We assume that the speed of light
is invariant and that the position \coo transform in such a way to ensure the
invariance of the deformed symplectic structure of phase space.

\vfill
\end{titlepage}

\section{Introduction}
Some years ago, the idea that special relativity should be modified in such a way
that the Planck energy $\k$ be an invariant parameter like the speed of light
was proposed \cite{AC1}. Models based on this idea were named doubly special relativity
(DSR).
Originally, the idea was implemented through a deformation of Lorentz transformations
acting on momentum space, which in turn implied the deformation of the dispersion relations
of elementary particles \cite{AC1}. In particular, it was assumed that only proper Lorentz
\trans were deformed, while rotations were realized canonically.
Several different models of this kind were proposed, based on different deformations
\cite{MS1,dsr}. The analysis performed in these papers was limited to the momentum sector
of phase space, while the spacetime realization of the deformations was not discussed.

This led to some criticism, based on the observation that nonlinear \trans of the momenta
could bring back the action of the Lorentz group to its canonical form \cite{Lu}, making
therefore the theory trivial.
Even accepting this point of view, it should however be noted that realizations of DSR
on position space give rise to predictions that are certainly not equivalent to those of
special relativity.

Later, in fact, different realizations on spacetime of the DSR idea were proposed.
First of all, it was realized that existing models based on $\k$-\poi algebras and
$\k$-\min spacetime \cite{LNR} fitted very well the DSR axioms.
Other classical realizations were mainly based on the definition of appropriate
transformation laws for the positions coordinates, which enforced the invariance of the
(not necessarily canonical) symplectic structure of phase space\footnote{This idea was
first outlined in \cite{LRR} in the context of $\k$-\min spacetime.} \cite{Gr,Mi1,Mi2,str}.
A similar proposal was also advanced in \cite{KM}. In this paper, however, the
transformation laws for the positions coordinates were not assumed to preserve the
symplectic structure, but rather the scalar product of positions with momenta.
A critical discussion of these models can be found in \cite{GM}.

It also appeared that the most natural realization of the DSR idea is in terms of
noncommutative spacetime. Recently, a realization of DSR on noncommutative spacetime
with no modifications of the dispersion relations has also been proposed \cite{AC2}.

In general, the phenomenological consequences of DSR depend on the spacetime model
adopted \cite{AK}, and especially on the definition of the velocity of a particle
\cite{vel,Mi3,DIK}.
In this letter we consider a specific proposal for the spacetime realization of DSR
\cite{Mi2,Mi1,Mi4}, based exclusively on classical particle
mechanics, that starting from any given momentum space realization of DSR, defines its
position space realization by requiring covariance of the phase space coordinates.
To distinguish this model from the numerous different realizations of DSR, we call it
canonical DSR.

This proposal is based on two requests: define the transformation law of position \coo in
such a way that they leave the symplectic form of phase space invariant,
and obtain a definition of velocity compatible with its identification as a parameter of
Lorentz transformations \cite{vel,Mi3,DIK}. The last demand enforces the deformation of the
canonical structure of phase space, yielding nonvanishing \pb between position coordinates
\cite{Mi1,DIK}, that can be interpreted as a classical counterpart of noncommutative geometry.
It is also important to notice that our definition of velocity is such that the speed of
light is really a constant, thus avoiding the problems related to a variable speed of light
present in some spacetime realizations of DSR \cite{KM}. In this letter, we shall not expose
in detail the formalism of canonical DSR, but refer to the above-cited papers
\cite{Mi2,Mi1,Mi4}.

The result of our analysis is that in the best known examples of DSR \cite{MS1,LNR},
canonical DSR implies nontrivial corrections to the time of flight of unstable particles
with respect to the predictions of special relativity. This possibility was already put
forward in ref.\ \cite{Mi4}. A discussion of time of flight of particles based on the
spacetime realization of DSR of ref.\ \cite{KM} is given in ref.\ \cite{Ho}. Due to the
considerable differences in the two approaches (in particular the variability of the speed of
light assumed in \cite{KM}), the results of \cite{Ho} do not apply here.

In the following we rise and lower indices with the flat metric of signature $(+,-,-,-)$.
Greek indices run from 0 to 3 and latin indices from 1 to 3.

\section{The Magueijo-Smolin model}
As a first example of application of the formalism, we consider the Magueijo-Smolin (MS) model
\cite{MS1}, that is the simplest realization of DSR in momentum space.
It can be characterized by the deformation of the transformation law of the momentum of a particle
under boosts. For a boost in the $x^1$ direction with rapidity parameter $\x$, the momentum
transforms as \cite{MS1}
\bea\lb{ptr}
&&p'_0={p_0\cosh\x+p_1\sinh\x\over\D(p_\m)},\qquad
p'_1={p_1\cosh\x+p_0\sinh\x\over\D(p_\m)},\cr
&&p'_2={p_2\over\D(p_\m)},\qquad\qquad p'_3={p_3\over\D(p_\m)},
\eea
where
\be\lb{del}
\D(p_\m)=1+{p_0(\cosh\x-1)+p_1\sinh\x\over\k}.
\ee

The dispersion relation of the MS model, invariant under the
transformations (\ref{ptr}) is
\be\lb{cas}
{p_0^2-p_i^2\over\den^2}=m^2,
\ee
where $m$ is the Casimir mass. This is related to the rest energy $M$
of the particle by
\be
m={M\over1-{M\over\k}}.
\ee
From (\ref{ptr}) one can derive a relation between the rapidity parameter
and the energy $p_0$ of a particle in a frame related to its rest frame by a
boost of parameter $\x$ \cite{Mi3},
\be\lb{mom}
p_0={M\g\over1+{M\over\k}(\g-1)},
\ee
where $\g\id\cosh\x$. Inverting, one obtains the parameter $\g$ as a
function of the energy $p_0$,
\be\lb{gam}
\g={p_0\over M}{\dem\over\den}.
\ee

Defining the 3-velocity $v_i$ of a particle in a suitable way, it is possible to
identify $\g$ with its classical expression, $\g=(1-v_i^2)^{-1/2}$ \cite{vel,Mi3}.
This condition is necessary if one requires that the transformation law of the
velocity of a particle under boosts be independent of its mass \cite{Mi3},
and permits to identify the velocity with the parameter of the Lorentz
transformations \cite{vel}. Moreover, it preserves the Einstein formula for the
composition of velocities and implies that the speed of light is independent of
the energy.
Such definition of velocity can be obtained by deforming the canonical symplectic
structure as \cite{Gr,Mi1,DIK}
\bea\lb{sym}
&&\{x^0,x^i\}={x^i\over\k},\quad\{p_0,p_i\}=0,\quad\{x^0,p_0\}=
1-{p_0\over\k},\cr
&&\{x^i,p_j\}=\d^i_j,\quad\{x^0,p_i\}=-{p_i\over\k},\qquad\{x^i,p_0\}=0,
\eea
and imposing that the position variables $x^\m$ transform in such a way that,
combined with (\ref{ptr}), leave invariant the corresponding symplectic structure
\cite{Mi1,Mi2}, namely\footnote{The same transformations were proposed in \cite{KM},
although starting from different assumptions.}
\bea\lb{xtr}
&&x'^0=\D(p_\m)(x^0\cosh\x-x^1\sinh\x),\qquad
x'^1=\D(p_\m)(-x^0\sinh\x+x^1\cosh\x),\cr
&&\cr
&&x'^2=\D(p_\m)\, x^2,\qquad\qquad x'^3=\D(p_\m)\, x^3,
\eea
with $\D(p_\m)$ given by (\ref{del}). We observe that, as usual in DSR theories, a
consistent
definition of the Lorentz \trans of the position \coo must be momentum dependent.

Choosing the Hamiltonian proportional to the Casimir invariant (\ref{cas}),
the Hamilton equations then read \cite{Gr,Mi1}
\be
\dot x_0={p_0\over\den^2},\qquad\dot x_i={p_i\over\den^2},
\ee
\be
\dot p_0=\dot p_i=0,
\ee
The classical definition of 3-velocity follows,
\be
v_i\id{\dot x_i\over\dot x_0}={p_i\over p_0}.
\ee
It is also easy to verify that the metric \cite{Gr,Mi1}
\be\lb{lin}
ds^2=\den^2d\bar s^2,
\ee
where $d\bar s^2$ is the Minkowski metric,
is invariant under the deformed transformations (\ref{ptr}), (\ref{xtr}).
A noticeable property of the metric (\ref{lin}) is its momentum dependence.
This also is a common feature of DSR models and has led to a proposal
for their generalization to include gravity \cite{MS3}. Note that in our formalism
the causal structure is not affected by the dependence on momenta.
The metric must be interpreted as that experienced by a particle of energy $p_0$.
A similar situation arises in scalar-tensor gravity, where particles with different
scalar coupling experience different metric structures. Using (\ref{mom}) one may
also interpret the metric as dependent on the mass and the velocity of the particles,
rather than momenta, obtaining a structure similar to, but not coincident with, that
of Finsler geometry \cite{fin}.

\bigskip
Let us now proceed to calculate the proper time of a particle with \coo
$x^\m$ in terms of the time measured by an observer of \coo $x'^\m$ at rest
in the laboratory.
For a small displacement $dx^\m$, using (\ref{xtr}) one has, since $dx^i=0$,
\be
dx'^0=\D(p_\m)\,\g\ dx^0,
\ee
where $\D(p_\m)$ is given by (\ref{del}) with $p_\m=(M,0,0,0)$.
Calling $t$ the laboratory time $x'^0$, and $\t$ the proper time $x^0$,
one has, more explicitly,
\be\lb{pt}
dt=\g\left[1+{M\over\k}(\g-1)\right]d\t.
\ee

The same result can be obtained from the invariance of the line element (\ref{lin}).
Equating (\ref{lin}) calculated in the laboratory and rest frames, it results
\be
\den^2(1-v_i^2)\,dt^2=\den^2\g^{-2}dt^2=\dem^2d\t^2.
\ee
Substituting (\ref{mom}), one recovers (\ref{pt}).

From a phenomenological point of view, it may be more useful to write
$dt$ as a function of the energy $p_0$ of the particle. Using (\ref{gam}), one
gets
\be
dt={p_0\over M}{\dem^2\over\den^2}\ d\t\approx{p_0\over M}
\left(1+2\,{p_0-M\over\k}\right)d\t.
\ee
It is then evident that the formalism of canonical DSR implies corrections to the
time of flight formula of special relativity of order $p_0/\k$. In particular,
an observer at rest in the laboratory measures a value of the lifetime of an unstable
particle greater than that predicted by special relativity. Although the corrections
are too small to be detected at present, they are in principle observable.

\section{The Lukierski-Nowicki-Ruegg model}
An analogous calculation can be done for the Lukierski-Nowicki-Ruegg (LNR) model
\cite{LNR}. This analysis has already been performed in ref.\ \cite{Mi4}, but
here we rederive that result starting from the invariance of the line element.

In the LNR model a deformed boost in the $x^1$ direction acts on the momentum variables
as \cite{BAK}
\bea\lb{trp}
&&p'_0=p_0+\k\log\G,\qquad p'_1={p_1\cosh\x+{\k\over2}\left(1-e^{-2p_0/\k}+
{p_k^2\over\k^2}\right)\sinh\x\over\G(p_\m)}\, ,\cr
&&p'_2={p_2\over\G(p_\m)},\qquad\qquad\qquad p'_3={p_3\over\G(p_\m)},
\eea
where
\be
\G(p_\m)=\ha\left(1+e^{-2p_0/\k}-{p_k^2\over\k^2}\right)+\ha\left(1-e^{-2p_0/\k}
+{p_k^2\over\k^2}\right)\cosh\x+{p_1\over\k}\sinh\x.
\ee

The dispersion relation invariant under the deformed boosts reads
\be\lb{disp}
4\k^2\sinh^2{p_0\over2\k}-p_k^2e^{p_0/\k}=m^2,
\ee
with $m$ the Casimir mass, related to the rest mass $M$ by
\be\lb{mas}
m=2\k\sinh{M\over2\k}.
\ee
The definition of velocity consistent with its identification as the parameter of
the Lorentz \trans is \cite{vel,DIK,LN}
\be\lb{velo}
v_i={2p_i\,e^{p_0/\k}\over\k\left(e^{p_0/\k}-\cosh{M\over\k}\right)},
\ee
and the parameter $\g=\cosh\x$ of the transformation between the laboratory and the
rest frames of a particle of energy $p_0$ is given by \cite{Mi3}
\be\lb{gamma}
\g={e^{p_0/\k}-\cosh{M\over\k}\over\sinh{M\over\k}}.
\ee

The definition (\ref{velo}) can be obtained by postulating the Poisson brackets
\cite{Mi4}
\bea\lb{symp}
&&\{x_0,x_i\}=2\,{p_ix_0\over\k^2}-\left(1+{p_k^2\over\k^2}\right){x_i\over\k},
\quad\{p_0,p_i\}=0,
\cr
&&\{x_0,p_0\}=\ha\left(1 + e^{-2p_0/\k}+{p_k^2\over\k^2}\right),
\quad\{x_0,p_i\}=-{p_i\over\k}e^{-2p_0/\k},\cr
&&\{x_i,p_0\}={p_i\over\k},\quad\quad\{x_i,p_j\}=-e^{-2p_0/\k}\d_{ij}.
\eea
Note that that these \pb are different from the standard ones adopted in $\k$-\min
space \cite{LNR}.

Given the symplectic structure (\ref{symp}), the \tran of the position \coo under
a boost in the $x^1$ direction, contravariant with respect to (\ref{trp}) are
\cite{Mi4}
\bea\lb{trx}
&&x'^0={x^0\cosh\x-x^1\sinh\x\over\G(p_\m)},\qquad
x'^1={x^1\cosh\x-x^0\sinh\x\over\G(p_\m)},\cr
&&x'^2={x^2\over\G(p_\m)}\,,\qquad\qquad x'^3={x^3\over\G(p_\m)}\,,
\eea
Choosing the Hamiltonian proportional to the Casimir invariant (\ref{disp}),
the Hamilton equations read
\bea
&&\dot x_0={\k\,e^{2p_0/\k}\over8m}\left(1+e^{-2p_0}-
{p_k^2\over\k^2}\right)^2\left(1-e^{-2p_0/\k}+{p_k^2\over\k^2}\right),\cr
&&\dot x_i={e^{2p_0/\k}\over4m}\left(1+e^{-2p_0}-{p_k^2\over\k^2}\right)^2p_i,
\eea
from which one can recover the velocity (\ref{velo}).
The metric invariant under (\ref{trp}) and (\ref{trx}) reads \cite{Mi4}
\be
ds^2={16\,e^{-2p_0/\k}\over(1+e^{-2p_0}-p_1^2/\k^2)^4}\ d\bar s^2.
\ee
For a particle of rest mass $M$ the metric can be written, using (\ref{disp}) and
(\ref{mas}),
\be\lb{met}
ds^2={e^{2p_0/\k}\over(2\cosh(M/\k)-1)^2}\ d\bar s^2.
\ee
Equating (\ref{met}) evaluated in the laboratory frame and in the rest frame, one gets,
using the same notations as in previous section,
\be\lb{dtt}
{e^{p_0/\k}\over2\cosh(M/\k)-1}\ {dt\over\g}={e^{M/\k}\over2\cosh(M/\k)-1}\ d\t.
\ee
On the other hand, the energy $p_0$ of the particle measured in the laboratory frame
can be written in terms of $\g$ as (cfr.\ (\ref{gamma}))
\be
e^{p_0/\k}=\cosh(M/\k)+\g\sinh(M/\k).
\ee
Substituting in (\ref{dtt}) one gets
\be
dt={2\g\over 1+\emm+\g(1-\emm)}\ d\t.
\ee
This is the result obtained in \cite{Mi4} by a different method.
Using (\ref{gamma}) one can also write this relation in terms of the energy $p_0$
as
\be
dt=2\ {1-\cosh{M\over\k}\,e^{-p_0/\k}\over1-e^{-2M/\k}}\ d\t\sim{p_0\over M}
\left(1-{p_0-M\over\k}\right)d\t.
\ee
Notice that in this case the correction has opposite sign with respect to that
obtained for the MS model, and hence the predicted lifetimes of unstable flying
particles are smaller than the ones of special relativity.

\section{Conclusions}
We have shown that the formalism of canonical DSR induces nontrivial corrections to the
time of flight of an unstable particle, with respect to the predictions of special
relativity. This shows the nontriviality of DSR when a spacetime realization is given.
The corrections depend on the specific model of DSR under consideration, and are related
to the deformation of the Lorentz invariance. It is easy to see in fact that models of
DSR that preserve the Lorentz invariance, like the
Snyder model \cite{Sn} do not exhibit such effect.

It must also be remarked that the results we have obtained depend strongly on the
spacetime realization of DSR, and in particular on the definition of velocity.
Different spacetime models will lead to different conclusions, see for example \cite{Ho}.
Unfortunately, there is no agreement yet on this topic in the literature, and the
eventual choice of a specific realization should be considered a matter of experiment.

On the other hand, an experimental check of our results seems out of reach at present.
The most favourable situation for measurements of quantum-gravity effects seems to be the
detection of neutrinos associated with gamma ray burst, whose energies are around $10^5$
GeV \cite{AP}.  In the atmosphere, they could produce muons or tau particles of similar
energies. Even if the lifetime of these particles could be
measured, the corrections to the relativistic formula would be of order $10^{-12}$, which
seems beyond observational power. The effects predicted could however be detectable if some
mechanism fixes the scale of $\k$ well below the Planck energy.

Finally, we remark that, although our realization of DSR solves the problems related to
the definition of velocity, it leaves other puzzles open, in particular the definition
of multiparticle states and the macroscopic limit of the theory. The
first problem could become relevant when applying our results to composite particles.

\end{document}